\documentclass{article}
\usepackage{spconf}

\usepackage{amsmath, bbm}
\usepackage{amssymb}
\usepackage{amsfonts}
\usepackage{amsthm}
\usepackage{algorithm}
\usepackage{algorithmic}
\usepackage{nicefrac}
\usepackage{bm}
\usepackage{hyperref}
\usepackage{cite}
\usepackage[utf8]{inputenc} 
\usepackage[T1]{fontenc}
\usepackage{graphicx}
\usepackage[font=small]{caption}
\usepackage{tikz}
\usetikzlibrary{positioning}
\usepackage{pgfplots}
\usepackage{epstopdf}
 \usepackage[labelformat=simple]{subcaption}
\usepackage{multirow}
\usepackage{lipsum}
\usepackage{mathtools}

\long\def\comment#1{}

\input{mysymbol.sty}

\title{Congestion-aware Distributed Task Offloading in Wireless Multi-hop Networks Using Graph Neural Networks}

\name{Zhongyuan Zhao$^\star$, Jake Perazzone$^\ddag$, Gunjan Verma$^\ddag$, 
and Santiago Segarra$^\star$
\thanks{Research was sponsored by the Army Research Office and was accomplished under Cooperative Agreement Number W911NF-24-2-0008 and W911NF-19-2-0269. 
The views and conclusions contained in this document are those of the authors and should not be interpreted as representing the official policies, either expressed or implied, of the Army Research Office or the U.S. Government. 
The U.S. Government is authorized to reproduce and distribute reprints for Government purposes notwithstanding any copyright notation herein.
\newline
Emails: $^\star$\{zhongyuan.zhao, segarra\}@rice.edu, $^\ddag$\{jake.b.perazzone.civ, gunjan.verma.civ\}@army.mil}}
\address{$^\star$Rice University, USA \hspace{10mm} $^\ddag$US Army’s DEVCOM Army Research Laboratory, USA}

\begin{document}
\ninept
\renewcommand{\baselinestretch}{0.98}
\maketitle
\begin{abstract}
% Computational offloading is becoming increasingly important to mobile and smart devices as these devices are tasked with artificial intelligence (AI) based algorithms.
Computational offloading has become an enabling component for edge intelligence in mobile and smart devices.
Existing offloading schemes mainly focus on mobile devices and servers, while ignoring the potential network congestion caused by tasks from multiple mobile devices, especially in wireless multi-hop networks. 
To fill this gap, we propose a low-overhead, congestion-aware distributed task offloading scheme by augmenting a distributed greedy framework with graph-based machine learning. % \textcolor{blue}{GV: should we add "low-overhead"?}
In simulated wireless multi-hop networks with 20-110 nodes and a resource allocation scheme based on shortest path routing and contention-based link scheduling, our approach is demonstrated to be effective in reducing congestion or unstable queues under the context-agnostic baseline, while improving the execution latency over local computing.
% \textcolor{blue}{GV: what exactly is "task congestion"? do we mean network congestion wherein multiple tasks share common links?} 
% \textcolor{blue}{GV: can we give actual numbers instead, e.g. "networks of size 10-100"?}
% at the cost of increased execution latency for tasks with rapid responses  \textcolor{blue}{GV: did not fully follow the meaning of this last phrase; do we mean "at the cost of a latency overhead (compared to the greedy heuristic) of the execution of our graph-based ML algorithm"?} .
\end{abstract}
\begin{keywords}
Computational offloading, queueing networks,
% \textcolor{blue}{GV: I vote for removing this keyword or perhaps changing it to "distributed scheduling" or something more informative} 
wireless multi-hop networks, graph neural networks, shortest path.
\end{keywords}

\vspace{-0.05in}
\section{Introduction}\label{sec:intro}
\vspace{-0.05in}

The proliferation of mobile and smart devices enables the collection of rich sensory data from both physical and cyber spaces, leading to many exciting applications, such as connected vehicles, drone/robot swarms, software-defined networks (SDN), and Internet-of-Things (IoT) \cite{sarkar2013ad,kreutz2014software,kott2016internet,cisco2020}.
To support these applications, wireless multi-hop networks, which have been traditionally used for military communications, disaster relief, and sensor networks, are now envisioned in the next-generation wireless networks, including xG (device-to-device, wireless backhaul, and non-terrestrial coverage), vehicle-to-everything (V2X), and machine-to-machine (M2M) communications~\cite{akyildiz20206g,chen2021massive,noor20226g}.
This can be partially attributed to the self-organizing capability of wireless multi-hop networks, enabled by distributed resource allocation without relying on infrastructure ~\cite{tassiulas1992,Joo09,Lin2009constant,jiang2010distCSMA,zhao2022twc,zhao2023delay,zhao2023graphbased}.
Additionally, computational offloading~\cite{van2018quality,ferrer2019towards,funai2019computational,chattopadhyay2020fully,cai2020mobile,feng2021multi,liu2022mobility,dai2022learning,li2022novel} promises to improve the performance and energy efficiency of resource-limited mobile devices by moving their resource-intensive computational tasks to external servers, including remote computing clusters and edge servers located near the mobile devices~\cite{ferrer2019towards}. 
% \textcolor{blue}{GV: small point, but "continuum" may be a stretch given there are a few discrete choices; how about "....remote computing clusters, resulting in a plethora of joint edge-fog-cloud offloading options". } 
In particular, computational offloading has become an enabling technology for edge intelligence,
as it is often impractical to equip mobile devices with hardware accelerators due to economic or energy constraints.
% \textcolor{blue}{GV: REWORDING THIS PHRASE AS FOLLOWS: as most mobile devices cannot directly be equipped with hardware accelerators for economic or energy reasons.} 

\begin{figure}
    \centering
    % \captionsetup[subfigure]{aboveskip=4pt,belowskip=4pt}
    \vspace{-0.2in}
    \hspace{-2mm}
    \subfloat[]{
    \includegraphics[height=1.0in]{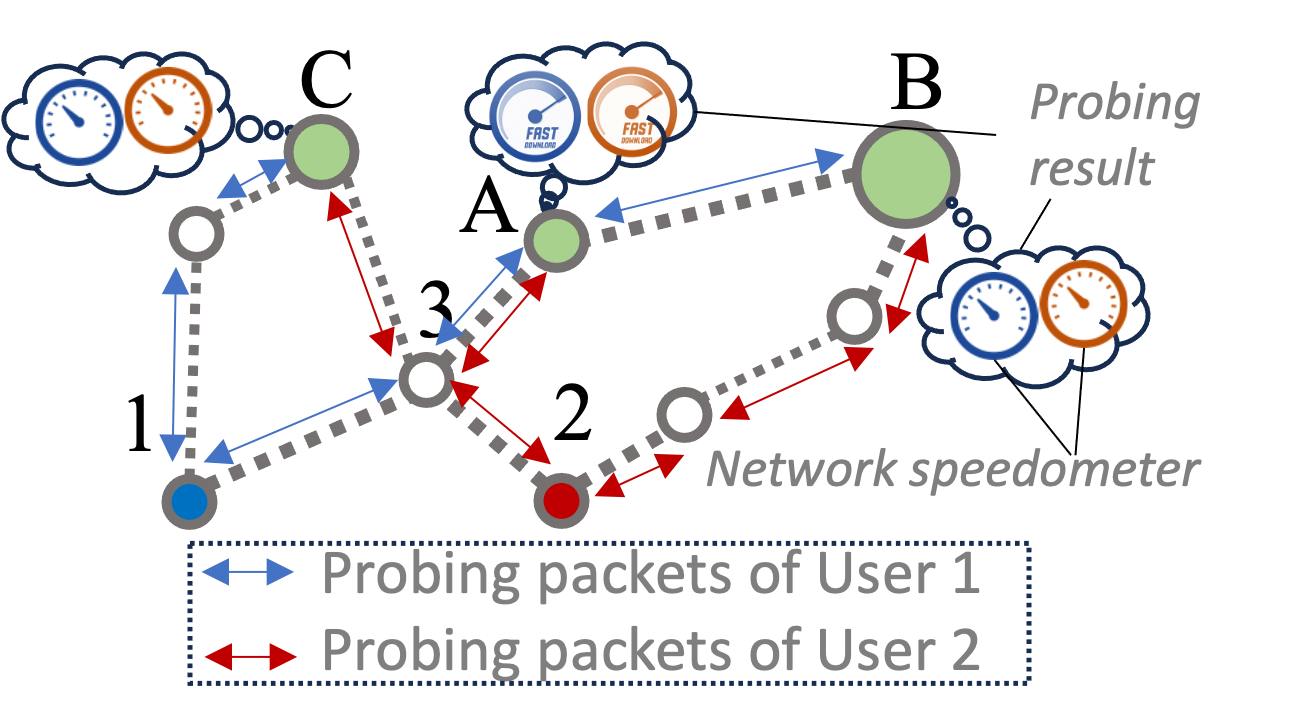}
    \label{fig:motivation:probe}\vspace{-0.1in}
    } 
    \subfloat[]{
    \includegraphics[height=1.0in]{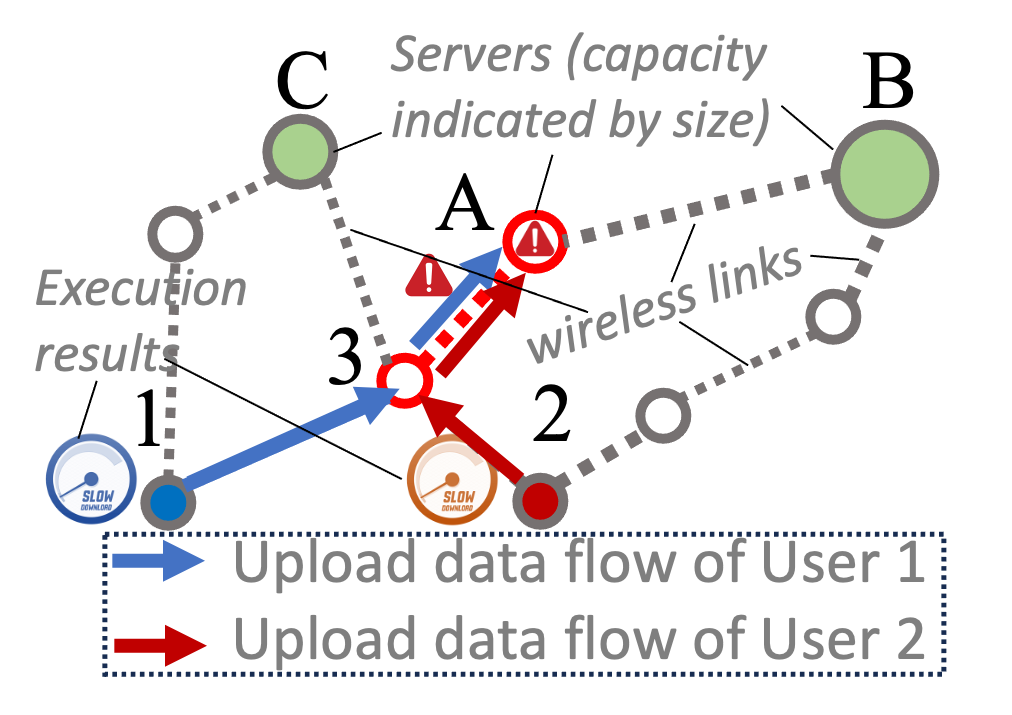}
    \label{fig:motivation:execution}\vspace{-0.1in}
    }
    \vspace{-0.1in}
    \caption{{\small Challenges in distributed multi-hop offloading:
    (a) probing: nodes $1$ and $2$ query the communication and computing bandwidth of three servers. 
    (b) offloading: nodes $1$ and $2$ both select server A based on collected information, however, such decisions lead to congestion at both the server A and link (3,A) in execution. % \textcolor{blue}{GV: can we explain what the "size" of the green nodes means? also, what exactly are the "clocks" indicating? what do blue and red arrows mean? what do dashed lines mean? we can add a legend instead of words, to save space.} 
    }}
    \label{fig:motivation}
    \vspace{-0.2in}
\end{figure}

Current studies in computational offloading (also referred as mobile edge computing, fog computing, cloudlets, etc.) mostly focus on the performance and energy consumption of individual devices~\cite{van2018quality,li2022novel,feng2021multi} under simplified networking assumptions, e.g., servers are within a single-hop~\cite{li2022novel,van2018quality}.
For offloading in wireless multi-hop networks~\cite{funai2019computational,chattopadhyay2020fully,cai2020mobile,feng2021multi,liu2022mobility,dai2022learning}, a centralized scheduler with global knowledge of the network is often assumed~\cite{funai2019computational,feng2021multi}.
However, centralized multi-hop offloading has the drawbacks of single-point-of-failure and poor scalability, due to the high communication overhead of collecting the full network state to a dedicated scheduler.
Distributed multi-hop offloading based on pricing   ~\cite{chattopadhyay2020fully,liu2022mobility} and learning~\cite{dai2022learning} only focus on the capacity of servers, while ignoring the potential network congestion caused by offloading~\cite{cai2020mobile}, as illustrated by the motivating example in Fig.~\ref{fig:motivation}.
In phase 1 (Fig.~\ref{fig:motivation:probe}), nodes $1$ and $2$ query the capacities of three servers and the corresponding links with probing packets, leading them to both conclude that server A offers the fastest execution.
In phase 2 (Fig.~\ref{fig:motivation:execution}), nodes $1$ and $2$ start offloading their tasks to server A via the blue and red routes decided in phase 1, causing congestion at link $(3,A)$ and server $A$ since their capacities cannot simultaneously support the traffic of the two tasks.
This problem is more pronounced for recurrent computational tasks, such as video surveillance and network traffic analysis. 
Moreover, the complexity of managing this issue by negotiation between devices or trial-and-error can increase exponentially with the number of mobile nodes. 
% \textcolor{blue}{GV: a bit of a devil's advocate (i.e. potential reviewer comment) position here: this issue would happen if the two tasks almost simultaneously made their decisions, which might be improbable. Second, adaptive routing schemes (such as those used in ad hoc networks) may re-route some traffic to avoid the bottleneck link. Third, I am guessing most offloading schemes dont just "offload and forget" but would have some adaptive behavior; here, perhaps $1$ and $2$ would notice that their observed latency was way higher than what their probe packet estimated, and could use some consensus mechanism to flip a coin and one of them could choose another server to offload to.} 

% To address the coupling between network conditions and offloading decisions, we develop an intelligent distributed task offloading scheme by augmenting distributed greedy offloading with graph-based machine learning.
In this work, we develop an intelligent distributed task offloading scheme that can exploit the network context via graph-based machine learning.
Specifically, we build a distributable graph neural network (GNN) that can encode the network topology and information from all links, servers, and tasks in the network into congestion-aware link weights.
Such link weights can mitigate network congestion (unstable queues) by enabling mobile devices to better estimate the costs of their offloading options in the presence of tasks on other mobile devices, leading to improved task execution latency.

The contributions of this paper are twofold:\\
1) To our best knowledge, we present the first low-complexity approach to integrate network context into fully distributed and near-simultaneous offloading decisions in wireless multi-hop networks.  \\
% \textcolor{blue}{GV: one might argue that simultaneous offloading decisions is a 0-probability event. Is it fair to say "nearly simultaneous" or something similar?} 
2) Our approach is proved to be able to mitigate congestion while offloading tasks in simulated wireless multi-hop networks. 
% \textcolor{blue}{GV: the last phrase, i.e. "with shortest path...." could be deleted IMO.} 

\vspace{-0.1in}
\section{System Model and Problem Formulation}
\label{sec:problem}
\vspace{-0.05in}

We model a wireless multi-hop network as a \emph{connectivity graph} $\ccalG^{n}$ and a \emph{conflict graph}~$\ccalG^{c}$. 
The connectivity graph is an undirected graph $\ccalG^{n}=(\ccalV, \ccalE)$, where $\ccalV$ is a set of nodes representing wireless devices in the network and $\ccalE$ is a set of links in which $e=(v_1,v_2)\in\ccalE$ for $v_1,v_2\in\ccalV$ indicates that nodes $v_1$ and $v_2$ can communicate directly.
$\ccalG^{n}$ is assumed to be a connected graph, i.e., two arbitrary nodes in the network can always reach each other.
The conflict graph, $\ccalG^c=(\ccalE,\ccalC)$, describes the conflict relationship between links and is defined as follows: each vertex $e\in\ccalE$ corresponds to a link in $\ccalG^{n}$ and each undirected edge $(e_1, e_2)\in\ccalC$ indicates a conflict between links $e_1, e_2\in\ccalE$ in $\ccalG^{n}$.
Two links could be in conflict due to either 1) \emph{interface conflict}, i.e., two links share a wireless device with only one wireless transceiver; or 2) \emph{wireless interference}, i.e., their incident devices are within a certain distance such that their simultaneous transmission will cause the outage probability to exceed a prescribed level~\cite{cheng2009complexity}.
In this paper, we assume the conflict graph $\mathcal{G}^{c}$ to be known, possibly by each link monitoring the wireless channel~\cite{zhao2022twc}, or through more sophisticated estimation as in~\cite{yang2016learning}. 

Based on the role of each wireless device, $\ccalV$ can be partitioned into three subsets: $\ccalM$ for edge nodes, $\ccalR$ for relay nodes, and $\ccalS$ for server nodes.
An edge node $v\in\ccalM$ is a sensor with limited computing capability and power supply, such as IoT sensors, SDN routers, phones, wearable devices, drones, or robots.
A relay node $v\in\ccalR$ is dedicated to communications, such as satellite, fixed, or mobile relay stations.
A server node $v\in\ccalS$ has richer computing resources and power supply than edge nodes, but may be located multiple hops away from the requesting edge nodes. 

In this study, we consider a simplified task offloading scenario, in which each edge node may generate a non-trivial computational task for processing sensory data.
A task $j_m$ is a series of similar jobs generated by an edge node $m\in\ccalM$ at certain rate, and each job must be individually processed.
% \textcolor{blue}{GV: it will be helpful to define "recurrent jobs" more formally, e.g. perhaps as "a series of inputs generated with some rate, each of which must be individually processed"}
In particular, task $j_m$ encompasses 
%\textcolor{blue}{GV: let's use a different word here, maybe "encompasses"} 
the information of the source $m$, the set of available external servers $\ccalS_m\subseteq \ccalS$, the job arrival rate $\lambda^{j}(m)$, and the numbers of upload and download data packets per job, respectively denoted as $\eta^{u}(j_m)$ and $ \eta^{d}(j_m)$, where $\eta(j_m) = \eta^{u}(j_m) + \eta^{d}(j_m)$.
A task $j_m$ can be executed locally at its source $m$ or remotely on an external server $v\in\ccalS_m$ by uploading the data to the server $v$, and if required, sending back the results.
% \textcolor{blue}{GV: here we are presuming the sensor needs the output, which often may not be the case; if this is an essential assumption, perhaps we can state it up front, else we can modify as "and downloading the output data back, if required"}, where $\ccalS_m$ is assumed to be known to $m$.
We define set $\ccalS_m^+=\{m\}\cup \ccalS_m$ as the action space of offloading task $j_m$. 
We assume that all jobs are atomic, $\eta^{u}(j_m) \gg \eta^{d}(j_m)$,  jobs of different tasks arrive independently, and the execution time of a job grows by its description length.
% \textcolor{blue}{GV: it may be useful to define what exactly we mean by "independent", since the raw data itself is not statistically independent (e.g. data across frames of video).Also, if the downstream processing is an RNN, it will have memory and depend on multiple jobs (e.g. video frames). Maybe we only need to say "atomic" and can drop "independent". Or perhaps I have misunderstood what a "job" is.}  . 
Finally, we denote $\ccalJ$ as the set of all tasks in the network.

We consider a time-slotted medium access control in the network.
We model each wireless link $e\in\ccalE$ as a $G/G/1$ queueing system with a first-in-first-out (FIFO) queue, a single server, and general packet arrival and service processes, where the arrival and service rates are $\lambda(e)\geq 0$ and $\mu(e)\geq 0$, respectively.
Under this model, a packet leaves the queueing system when it reaches the other end of the link.
Based on Little's law~\cite{little1961proof}, if $\mu(e) > \lambda(e)$, the probability of a link having a non-empty queue is $\lambda(e)/\mu(e)$ and the expected time of a packet passing through a link is the response time of the queueing system, $1/(\mu(e)-\lambda(e))$.
If $\mu(e) \leq \lambda(e)$, the queue becomes unstable and will grow infinitely, which is referred as congestion. 
However, to quantify the level of congestion, we use the expected time to deplete the queue of a link, $T\lambda(e)/\mu(e)$, assuming that new jobs only arrive in the first $T$ time slots. 
% \textcolor{blue}{GV: why can we assume this?}.
Notice that $\mu(e)$ of a link is generally unknown as it depends on both the average link rate $r(e)$  and the link scheduling policy.
A similar queueing model also applies to a computing node $v\in\ccalM\cup\ccalS$, i.e., an edge or server node, except that the service rate $\mu(v)$ is known in advance and not affected by link scheduling.
Here, the arrival rate, link rate, and service rate are all defined in terms of number of packets per time slot.

We denote the features of links, nodes, and tasks under the following convention unless otherwise specified.
For example, vector $\bbr^c=\left[r(e)|e\in\ccalE\right]$ collects the link rates of all the wireless links, where superscript $c$ indicates that the dimension of $\bbr^c$ equals the number of nodes in graph $\ccalG^c$, i.e., $\bbr^c\in\reals^{|\ccalE|}$. % \textcolor{blue}{GV: we mean superscript, right?} 
Similarly, vector $\bbmu^n\in\reals^{|\ccalV|}$ describes the service rates of the node set $\ccalV$ in graph $\ccalG^n$ and
vector $\bbu^j\in\reals^{|\ccalJ|}$ represents the execution latency (in time slots) of all tasks. %\textcolor{blue}{GV: how is this quantity defined?}  
Matrices are denoted by upright bold upper-case symbols, where $\bbX_{ab}$ is the element at row $a$ and column $b$ of matrix $\bbX$, and $\bbX_{a*}$ is the $a$th row vector and $\bbX_{*b}$ is the $b$th column vector.

% We study the multi-hop offloading problem of deciding at which node $s_m\in\ccalS_m^+$ should a task $j_m\in\ccalJ$ from an edge node $m\in\ccalM$ be executed, with the objective of minimizing the expected execution latency of task $j_m$, denoted by $u(j, m, s_m)$.
Formally, the latency-optimal multi-hop offloading problem is formulated as finding the optimal offload locations $\bbs^{j*}$ to minimize the total expected execution latency across all tasks, $\boldsymbol{1}^{\top}\bbu^j$,
% The optimal multi-hop offloading can be formulated as follows:
\begin{subequations}\label{E:formulation}
\begin{align}
\bbs^{j*} &= \argmin_{\bbs^j} \sum_{j_m\in\ccalJ} u(j_m, m, s_m) \label{E:formulation:obj}\\
\text{s.t. } 
&  \bbs^j \coloneqq \left[s_m | j_m\in\ccalJ \right] ,\;\bbu^j \coloneqq \left[u(j_m, m, s_m)| j_m\in\ccalJ\right] \;, \label{E:formulation:vectors}  \\
& s_m\in\ccalS_m^+,   \forall\; m\in\ccalM \;, \label{E:formulation:action}\\
& \bbu^j = f_r\left(\ccalG^n,\ccalG^c, \bbmu^n, \bbr^c, \ccalJ, \bbs^j \right) \;, \label{E:formulation:utility}
\end{align} 
\end{subequations}
where $u(j_m, m, s_m)\geq0$ is the expected latency for task $j_m$ being executed on node $s_m$,
% \red{SS: missing subscript in $j$?}
% \textcolor{blue}{GV: as commented earlier, many instantiations of the offloading problem may not require a reply back to the sensor which originated the data; can we just say "end-to-end latency" or "task latency" or something more general than "response latency", allowing us to generalize to these scenarios as well?} 
and $f_r(\cdot)$ is a deterministic function that maps the network state $(\ccalG^n,\ccalG^c, \bbmu^n, \bbr^c, \ccalJ)$ and decision variables $\bbs^j$ to the expected execution latency of all tasks $ \bbu^j $, defined in \eqref{E:formulation:vectors}.
$f_r(\cdot)$ captures the effect of the resource allocation policy of the wireless network, such as the routing and scheduling protocols, on the expected execution latency of all the tasks under given offloading decisions $\bbs^j$.
% such as the routing and scheduling protocols that determine the round-trip route of a task $j_m$ between its source $m$ and server $s_m$, and the expected latency on each link and server for a job in task $j_m$.
Problem~\eqref{E:formulation} belongs to the class of generalized assignment problems (GAPs)~\cite{oncan2007survey}, which is known to be NP-hard.
% \textcolor{blue}{GV: we can define this acronym since it is used for the first time here}

% \textcolor{blue}{GV: maybe instead of "simplify" we can say "make concrete" or "instantiate", since our choices are not really "simple" in any sense, i.e. they are realistic choices. Another option is to just say "We specify the constraint as...." and later say "Note our approach is general and can easily be adapted to other scheduling and routing policies".} 
We specify the constraint in~\eqref{E:formulation:utility} with the following resource allocation schemes commonly found in practice: 
1)~a contention-based scheduling policy that offers conflicting links (neighboring nodes on the conflict graph) with non-empty queues an equal chance of transmission, e.g., CSMA~\cite{jiang2010distCSMA};
and 2)~a shortest path routing scheme that determines the route between source $m$ and an external server $s\in\ccalS_m$ for task $j_m$, based on the link weights; an example of such weights could be the expected time a data packet takes to pass through a link under the current traffic condition. 
% \textcolor{blue}{GV: I think the "e.g." here refers to the link weights, but as written this is not clear what the "e.g." refers to. we can instead reword to.....based on the link weights; an example of such weights could be the expected time....} 
Note that our approach can be easily adapted to other resource allocation schemes.

\vspace{-0.05in}
\section{Distributed Task Offloading with GNNs}
\label{sec:solution}
\vspace{-0.05in}

We propose to approximately solve Problem~\eqref{E:formulation} by augmenting distributed greedy decision-making with a trainable GNN.
% \textcolor{blue}{GV: approximate the problem, or its solution?} 
Our distributed decision framework is based on an extended connectivity graph, $\ccalG^e = (\ccalV^e, \ccalE^e)$, built by adding virtual nodes and links to the original connectivity graph $\ccalG^n$ as $\ccalV^e=\ccalV\cup\tilde{\ccalV}$ and $\ccalE^e=\ccalE\cup\tilde{\ccalE}$.
For each edge or server node $v\in\ccalM\cup\ccalS$, there is a corresponding virtual node $\tilde{v}\in\tilde{\ccalV}$ connected to $v$ by a virtual link $(v,\tilde{v})\in\tilde{\ccalE}$.
The link rate of a virtual link equals to the service rate of node $v$, i.e., $r\left({(v,\tilde{v})}\right) = \mu(v)$, and the link rate of a physical link remains the same.
We further introduce the line graph of $\ccalG^e$ as $\ccalG^{\ell}$:  each vertex in $\ccalG^{\ell}$ corresponds to an edge in $\ccalG^e$, and an edge in $\ccalG^{\ell}$ indicates that the two corresponding edges in $\ccalG^e$ share a common vertex~\cite{harary1960some}.
The vector of the extended link rates $\bbr^{\ell}\in\reals^{|\ccalE^e|}$ thus captures both the original link rates $\bbr^c$ and original service rates $\bbmu^n$.

The distributed task offloading decision, denoted as ${\bbs}^j =h(\ccalG^{\ell},\bbdelta^{\ell},\ccalJ) $, % \red{SS: The superscript in the graph should be $\ell$ instead of $l$?} Yes, it is changed.
lets each edge node select the offloading location of minimal cost based on given weights of the extended links $ \bbdelta^{\ell} $, 
\vspace{-0.05in}
\begin{subequations}\label{E:offload}
\begin{align}
& {\bbs}^j = \left[{s}_m | j_m\in\ccalJ \right],\;\; \text{where }\; {s}_m=\argmin_{v\in {\ccalS}_{m}^+} c(m,v) \;, \label{E:offload:decision}\\ 
& c(\!m,\!v\!) \!=\! \max\!\left[\eta^u(j_m)\beta(m,\!\tilde{v}) \!+\! \eta^d(j_m)\beta(v,\!m), 2\zeta(v,\!m)\right]\!.\label{E:offload:cost}
\vspace{-0.1in}
\end{align} 
\end{subequations}
In \eqref{E:offload}, the cost of offloading action $v\in {\ccalS}_{m}^+$, denoted as $c(m,v)$, is the expected round-trip delay of a job on graph $\ccalG^e$, while $\beta(v_1,v_2)$ and $\zeta(v_1,v_2)$ are the shortest path distance and hop distance between nodes $v_1, v_2$ on the edge-weighted graph $(\ccalV^e,\ccalE^e,\bbdelta^{\ell})$, respectively.
Eq. \eqref{E:offload:cost} says that a job will take at least one time slot to travel through a physical link, e.g., even if all the data packets of a job can go across a link within one time slot, they can only be sent over the next link until the next time slot.  
% \textcolor{blue}{GV: I did not find this part obvious, please explain. Also, have we formally defined the notion of "time slot" yet or that we assumed slotted time? If not, we can say this earlier.}

Next, we need to find a link weight vector, $\bbdelta^{\ell}= \left[\delta(e) | e\in \ccalE^e\right]$, that can be translated to good offloading decisions by the distributed decision framework $h(\cdot)$ in \eqref{E:offload}. 
% measure the expected per-packet delay on all extended links given the offloading decisions, which however are impossible to obtain due to the huge decision space, e.g., $\Pi_{m\in\ccalM}|\ccalS_m^+| $.
The baseline approach is to use the per-packet delay under a contention-free assumption,
% \textcolor{blue}{GV: would it be more precise to say ", in the absence of exogenous traffic, " instead of "the traffic-free"...since the offloading node of interest is introducing its own traffic?} 
e.g., $\bar{\bbdelta}^{\ell}=\boldsymbol{1}/\bbr^{\ell} $, 
which, however, only holds for scenarios with only a few tasks and lightly-loaded task traffic.

We propose to use a distributable GNN to predict a congestion-aware edge weight vector as $\hat{\bbdelta}^{\ell}=f(\ccalG^{\ell}, \bblambda^{\ell}, \bbr^{\ell}, \ccalG^c;\bbomega) $, where $\bbomega$ is the collection of trainable parameters of the GNN, and $\bblambda^{\ell}=\left[\lambda(e) |e\in\ccalE^e\right]$ assigns the packet arrival rates of tasks to corresponding virtual links, e.g., $\lambda(e)=\lambda(v)=\lambda^j(v) \eta(j_v)$ if $e=(v,\tilde{v})\in\tilde{\ccalE}$ and $j_v\in\ccalJ$, otherwise, $\lambda(v) = 0$.
% The GNN contains an  followed by another $K$ layers of non-trainable message passing operations.
In step 1, the GNN predicts the packet arrival rates on the extended links $\bbx^{\ell}\in\reals^{|\ccalE^e|}$, as $\bbx^{\ell} = \Psi_{\ccalG^{\ell}}(\bbX^0;\mathbf{\bbomega})$, where $\Psi_{\ccalG^{\ell}}$ is an $L$-layered graph convolutional neural network (GCNN) defined on graph $\ccalG^{\ell}$, and $\bbomega$ is the collection of trainable parameters.
We define the output of an intermediate $l$-th layer of the GCNN as $\bbX^l \in\reals^{|\ccalE^e|\times g_{l}}$, and the input and output dimensions of the GCNN are set as $g_{0}=4$ and $g_{L}=1$. 
The input node features are defined as $\bbX^0=\left[\bbq^{\ell},\bbw^{\ell},\bblambda^{\ell},\bbr^{\ell}\right]$,
where $ \bbq^{\ell} $ is an indicator vector of virtual links, i.e., $ \bbq^{\ell} = \left[ \mathbbm{1}_{\tilde{\ccalE}}(i) | i\in\ccalE^e \right] $, and $ \bbw^{\ell} $ is an indicator vector of virtual links for server nodes.
% of which the $i$th row encloses the features of edge $i=(v_1,v_2)\in\ccalE^e$, e.g., if edge $i$ is a virtual edge ($ v_2=\tilde{v}_1 $), then $\bbq^{\ell}_i=1$, $\bblambda^{\ell}_i =\lambda(v_1)$, and $\bbr^{\ell}_i=\mu(v_1)$, otherwise $\bbq^{\ell}_i=0$, $\bblambda^{\ell}_i =0$, and $\bbr^{\ell}_i=r(i)$ for $i\in\ccalE$.  

The $l$-th layer of the GCNN $\Psi_{\ccalG^{\ell}}$ can be implemented in a fully distributed manner through neighborhood aggregation as the following local operation on an extended link $i\in\ccalE^e$ (a vertex in $ \ccalG^{\ell} $)
\vspace{-0.08in}
\begin{equation}\label{E:gcn:local}
    \bbX_{i*}^{l} = \sigma_l \!\left(\!\bbX_{i*}^{l-1}  \bbTheta_{0}^{l} \! + \! \left[ \bbX_{i*}^{l-1} \!- \!\!\!\sum_{e \in \mathcal{N}^{\ell}(i)}\!\!\frac{\bbX_{i*}^{l-1}}{\sqrt{d^{\ell}({e})d^{\ell}({i})}} \right]\!\!\bbTheta_{1}^{l} \!\right),
\vspace{-0.08in}
\end{equation}
where $\bbX_{i*}^{l}\in\reals^{1\times g_{l}}$ captures the $l$th-layer features on link $i$, $\mathcal{N}^{\ell}(i)$ denotes the set of neighbors of $i$ in $\ccalG^{\ell}$, $d^{\ell}(\cdot)$ is the degree of a vertex in $\ccalG^{\ell}$, ${\bbTheta}_{0}^{l}, {\bbTheta}_{1}^{l} \in \mathbb{R}^{g_{l-1} \times g_{l}}$ are trainable parameters (collected in $\bbomega$), and $\sigma_l(\cdot)$ is an element-wise activation function of the $l$-th layer.
% \textcolor{blue}{COMMENT: we could put interfering in () here, since every neighbor in the conflict graph is an interferer}
Based on~\eqref{E:gcn:local}, the link arrival rate vector $\bbx^{\ell}$ can be computed in a fully distributed manner through $L$ rounds of local message exchanges between $i\in\ccalE^e$ and its neighbors on $\ccalG^{\ell}$.
% \red{SS: superscript $\ell$ missing?} 

% the service rates of the physical links $\bbmu^c=\left[ \mu(e) | e\in\ccalE\right]$ through link scheduling algorithm $K$ iterations of neighborhood aggregation, where the $k$th iteration $k\in\{1\dots K\}$ is, 

\begin{algorithm}[t]
\caption{{Iterative execution latency estimation algorithm $\varphi(\cdot)$}}
\label{algo:response}
\hspace*{\algorithmicindent} \textbf{Input}: $\ccalG^c, \ccalG^{\ell}, \bbr^{\ell}, \bbx^{\ell}, T, K$ \\
\hspace*{\algorithmicindent} \textbf{Output}: $ {\bbdelta}^{\ell} $ 
\begin{algorithmic}[1] 
\STATE Get $\bbA^c$ as the adjacency matrix of $\ccalG^c$
\STATE Get conflict degree vector $\bbd^c=\bbA^c\boldsymbol{1}^c$ 
\STATE $\bbx^c=\left[ \bbx^{\ell}_e | e\in\ccalE\right]$, $\bbmu^{c}(0) = \bbr^c/(\boldsymbol{1}^c + \bbd^c)$ 
\FOR{$ k \in \{1,\dots, K\} $ }
\STATE $ \bbb^{c}(k) = \min\left[{\bbx^c}/{\bbmu^{c}(k-1)},\boldsymbol{1} \right] $
\STATE $ \bbp^{c}(k) = \bbb^{c}(k)\bbA^c $
\STATE $ \bbmu^{c}(k) = {\bbr^c}/\left[\boldsymbol{1}^c+\bbp^{c}(k)\right]  $ 
\ENDFOR 
\STATE $\hat{\bbmu}^{\ell} = \left[ \hat{\mu}(e)|e\in\ccalE^e \right] $, where $ \hat{\mu}(e) = \bbmu^{c}_e(K) $ for a physical link $e\in\ccalE$ and $\hat{\mu}(e) =\bbr^{\ell}_{e}$ for a virtual link $e\in\tilde{\ccalE}$.
\STATE ${\bbdelta}^{\ell} = \left[ \delta(e)|e\in\ccalE^e \right] $, where $ \delta(e)=(\hat{\bbmu}^{\ell}_e-\bbx^{\ell}_e)^{-1}$ if $ \hat{\bbmu}^{\ell}_e > \bbx^{\ell}_e $,  otherwise $\delta(e)=T\bbx^{\ell}_e/\hat{\bbmu}^{\ell}_e $
\end{algorithmic}
\end{algorithm}
% \vspace{-0.2in}

In step 2, the congestion-aware weights for the extended links are estimated as $ \hat{\bbdelta}^{\ell} = \varphi(\ccalG^c, \ccalG^{\ell}, \bbr^{\ell}, \bbx^{\ell}, T,K) $, where function $\varphi(\cdot)$ is described by Algorithm~\ref{algo:response} and can be implemented in a fully distributed manner.
% ,  via $K$ rounds of local message exchanges on $\ccalG^c$.
In Algorithm~\ref{algo:response}, the service rate of each physical link, $\bbmu^c_e$, is initialized to the worst-case scenario that every neighboring links always contend for transmission.
Then, for each link $e\in\ccalE$, the algorithm iteratively updates: $ \bbb^c_e $, the probability of link $e$ contending for transmission due to non-empty queues, based on its input packet arrival rate $\bbx^c_e$ and service rate; $\bbp^c_e$, the expected number of contending neighbors of link $e$; and $\bbmu^c_e$ the service rate of link $e$ based on the scheduling policy that contending links have the same chance of transmission.
Lastly, the per-packet latency of the extended links are updated based on their arrival and service rates. 
% The iteration typically converges after $5$ iterations.

\vspace{1mm}
\noindent
{\bf Empirical task execution latency.}
Based on Algorithm~\ref{algo:response}, the empirical per-packet latency of the extended links can also be estimated as $ \bbtau^{\ell} = \varphi(\ccalG^c, \ccalG^{\ell}, \bbr^{\ell}, \bbrho^{\ell}, T, K)$. 
% \footnote{This method may not apply to trivial scenarios of ultra-light weighted traffic, e.g., the arrival rate and data size per job are too small such that the communication overhead dominates the execution latency.}.
% $ \left[ \rho(e) | e\in\ccalE^e\right] =  $
Here, vector $ \bbrho^{\ell}= \bbGamma \bblambda^{j} $ captures the traffic intensities on the extended links, which is determined by the packet arrival rates of all tasks $\bblambda^j=[\lambda(v)| (v,\tilde{v})\in\tilde{\ccalE}\And j_v\in\ccalJ ]$, 
and the uploading route indicator matrix $\bbGamma \in\{0,1\}^{|\ccalE^e|\times |\ccalJ|} $.
$\bbGamma$ is defined as: $\bbGamma_{e,j_m}=1 $ if link $e\in\ccalE^e$ is on the route from $m$ to the virtual node $\tilde{s}_m$ of the corresponding offloading action $s_m={\bbs}^j_m$ of task $j_m$, otherwise $\bbGamma_{e,j_m}=0 $.
The downloading route matrix $ \bbGamma^- $ is define as $ \bbGamma^-_{e*} = \bbGamma_{e*} \, \forall\;e\in\ccalE $, and $ \bbGamma^-_{e*} = \boldsymbol{0}^{\top} \, \forall\;e\notin\ccalE $. 
The empirical execution latency $ \bbu^j $ can be found as
\vspace{-0.05in}
\begin{equation}\label{E:latency}
    \bbu^j = f_u(\bbs^j) = \max\left[\bbtau^{\ell\top}\bbGamma \odot\bbeta^u + \bbtau^{\ell\top}\bbGamma^-\odot\bbeta^d,\; 2\boldsymbol{1}^{\top}\bbGamma^- \right].
\vspace{-0.05in}
\end{equation}
The empirical task execution latency vectors under the baseline and GNN-based policies  are $ \bar{\bbu}^j = f_u(\bar{\bbs}^{\ell}) $ and $ \hat{\bbu}^j = f_u(\hat{\bbs}^{\ell}) $, respectively, where 
$\bar{\bbs}^j =h(\ccalG^{\ell},\bar{\bbdelta}^{\ell},\ccalJ) $ and
$\hat{\bbs}^j =h(\ccalG^{\ell},\hat{\bbdelta}^{\ell},\ccalJ) $.

% \begin{subequations}\label{E:mu}
% \begin{align}
% \bbb^{c}(k) &= \min\{{\bbx^c}/{\bbmu^{c}(k-1)},\boldsymbol{1} \}\;, \label{E:mu:busy}\\
% \bbp^{c}(k) &= \bbb^{c}(k)\bbA^c \;, \label{E:mu:degree}\\
% \bbmu^{c}(k) &= {\bbr^c}/\left[\boldsymbol{1}+\bbp^{c}(k)\right] \;,\label{E:mu:rate}
% \end{align} 
% \end{subequations}
% where $\bbmu^{c}(0) = \bbr^c/(\boldsymbol{1} + \bbd^c)$, vectors $\bbd^c=\left[ d^c(e) | e\in\ccalE\right]$ and $\bbx^c=\left[ \bbx^{\ell}_e | e\in\ccalE\right]$ respectively collect the conflict degrees and predicted arrival rates of all physical links, and $\bbA^c$ is the adjacency matrix of $\ccalG^c$.
% The estimated service rate vector for the extended links is  $\hat{\bbmu}^{\ell} = \left[ \hat{\mu}(e)|e\in\ccalE^e \right] $, where $ \hat{\mu}(e) = \bbmu^{c}_e(K) $ for a physical link $e\in\ccalE$ and $\hat{\mu}(e) =\bbr^{\ell}_{e}$ for a virtual link $e\in\tilde{\ccalE}$.
% In each iteration in~\eqref{E:mu}, \eqref{E:mu:degree} involves neighborhood aggregation whereas \eqref{E:mu:busy} and \eqref{E:mu:rate} are just element-wise operations.
% Finally, the congestion-aware edge weight vector is computed as $\hat{\bbdelta}^{\ell} = \boldsymbol{1}/{(\hat{\bbmu}^{\ell}-\bbx^{\ell})} $. 

\begin{figure*}[t]
\centering
 %\vspace{-0.1in}
\hspace{-2.5mm}
\subfloat[]{
    \includegraphics[height=1.74in]{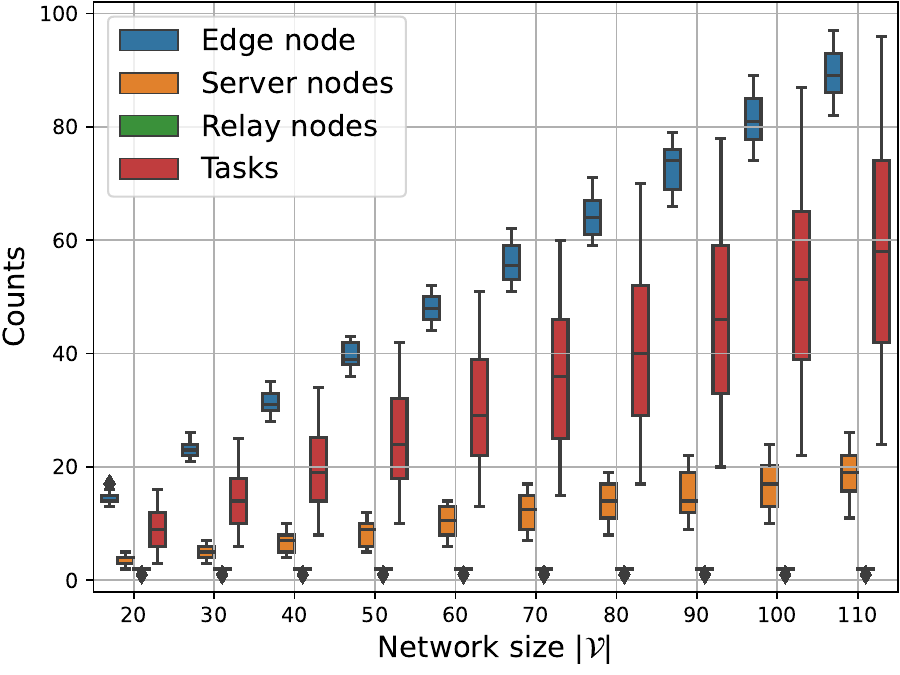}
    \label{fig:results:set}\vspace{-0.1in}
}% \hspace{-2mm}
\subfloat[]{
    \includegraphics[height=1.74in]{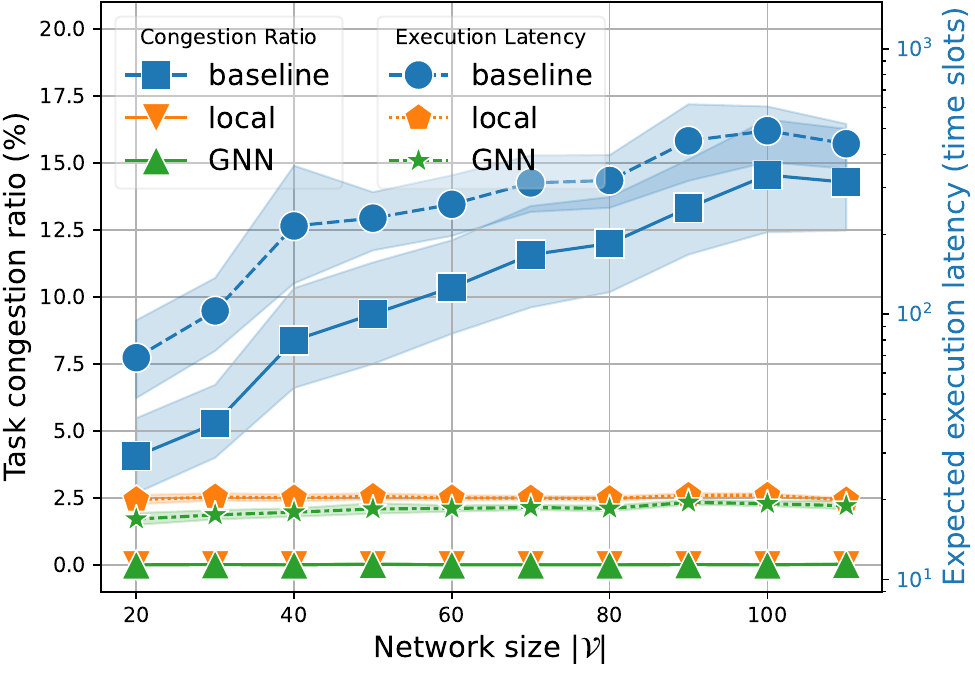}
    \label{fig:results:congest}\vspace{-0.1in}
}% \hspace{-2mm}
\subfloat[]{
    \includegraphics[height=1.74in]{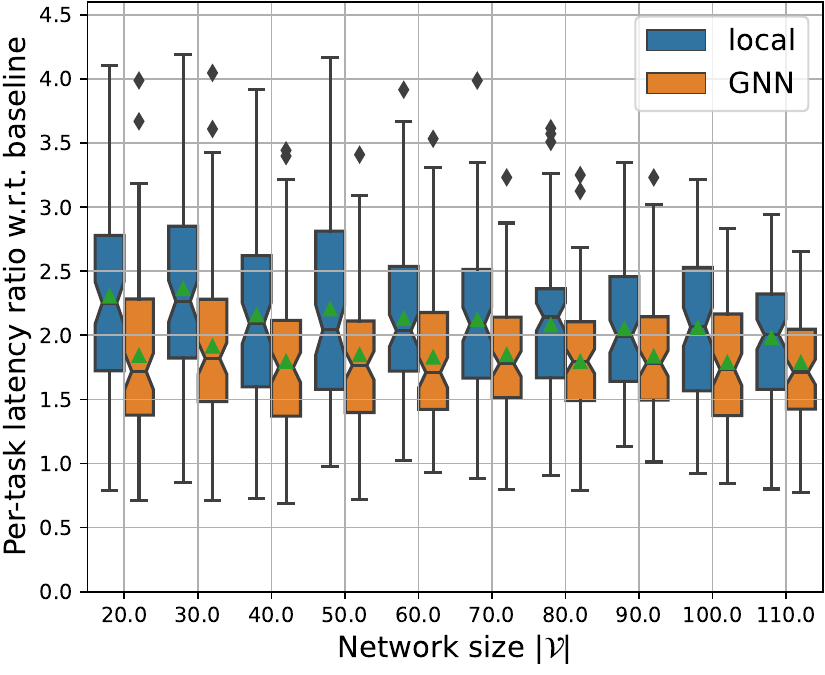}
    \label{fig:results:response}\vspace{-0.1in}
}
    \vspace{-0.1in}
    \caption{{\small 
    (a)~Scale of instances: the numbers of edge nodes, server nodes, relay nodes, and tasks by network size.
    (b)~Task congestion probability by network size with $95\%$ confidence interval. 
    (c)~Average per-task execution latency ratio w.r.t. the baseline policy across network sizes.
    }
    } 
 \label{fig:results}    
 \vspace{-0.2in}
\end{figure*}

\vspace{1mm}
\noindent
{\bf Complexity.}
For distributed execution, the local communication complexity (defined as the rounds of local exchanges between a node and its neighbors) of the GNN is $\ccalO(L+K)$.
% Hence, the local computational and communication costs can be controlled by modifying the number of layers $L$ in the GNN.
% The worst-case complexity of Dijkstra's algorithm is $\ccalO(|\ccalV|^2)$ \cite{dijkstra1959note}.
For the distributed decision framework in~\eqref{E:offload}, the distributed weighted single-source-shortest-path (SSSP) with the Bellman-Ford algorithm \cite{bellman1958routing,ford1956network} and all-pairs-shortest-path (APSP) with a very recent algorithm in~\cite{bernstein2019distributed} both take $\ccalO(|\ccalV|)$ rounds of message exchanges.

\vspace{1mm}
\noindent
{\bf Training.}
The parameters $\bbomega$ (collecting ${\bbTheta}_{0}^{l}$  and ${\bbTheta}_{1}^{l}$ across all layers $l$) of our GNN are trained on a set of random instances drawn from a target distribution $\Omega$.
Based on each instance of the form $\varepsilon=(\ccalG^{n}, \ccalG^{c}, \ccalM, \ccalR, \ccalS, \ccalJ, \bbr^c, \bbmu^n )\sim \Omega$, we create the corresponding extended graph $\ccalG^e$, its line graph $\ccalG^{\ell} $, $ \bblambda^{\ell}$, and $\bbr^{\ell} $, and run the full pipeline to get $\hat{\bbu}^j$. 
Since $\varphi(\cdot)$ and \eqref{E:latency} are differentiable, we can estimate the gradient of the objective $o(\bbomega)=\boldsymbol{1}^{\top}\hat{\bbu}^j$ in~\eqref{E:formulation:obj} w.r.t. $ \bbGamma $,
\vspace{-0.05in}
\begin{equation}\label{E:grad:route}
    \nabla_{\bbGamma} o(\bbomega)=\mathbb{E}_{\varepsilon\sim\Omega}\left[
     \frac{\partial \boldsymbol{1}^{\top}\hat{\bbu}^j(\varepsilon)}{\partial \bbGamma} \right],\;  \widehat{\nabla_{\bbGamma} o(\bbomega)}=
     \frac{\partial \boldsymbol{1}^{\top}\hat{\bbu}^j(\varepsilon)}{\partial \bbGamma}   \;.
\vspace{-0.05in}
\end{equation}
% \begin{equation}\label{E:grad:route}
%     \nabla_{\bbGamma} o(\bbomega)=\mathbb{E}_{\varepsilon\sim\Omega}\left[
%      \frac{\partial \boldsymbol{1}^{\top}\hat{\bbu}^j(\varepsilon)}{\partial \bbGamma} \right],\;  \widehat{\nabla_{\bbGamma} o(\bbomega)}=
%      \frac{\partial \boldsymbol{1}^{\top}\hat{\bbu}^j(\varepsilon)}{\partial \bbGamma}   \;.
% \end{equation}
We then estimate the gradient of $o(\bbomega)$ w.r.t. $\hat{\bbdelta}^{\ell}$ and $\bbomega$ as
% \textcolor{blue}{GV: gradient of, or gradient w.r.t. ? } 
\begin{subequations}
\vspace{-0.05in}
\begin{align}
    \widehat{\nabla_{\hat{\bbdelta}^{\ell}} o(\bbomega)} &= -\boldsymbol{1}^{\top}\widehat{\nabla_{\bbGamma} o(\bbomega)} + \nabla_{\hat{\bbdelta}^{\ell}} \left(\hat{\bbtau}^{\ell} - \hat{\bbdelta}^{\ell} \right)^2/|\ccalV^e| \;,\label{E:grad:delta}\\
    \widehat{\nabla_{\bbomega} o(\bbomega)} &= \widehat{\nabla_{\hat{\bbdelta}^{\ell}} o(\bbomega)} \left[{\partial f(\ccalG^{\ell}, \bblambda^{\ell}, \bbr^{\ell}, \ccalG^c;\bbomega)}/{\partial \bbomega}\right]\;. \label{E:grad:omega}
\vspace{-0.05in}
\end{align}
\end{subequations}
The estimation in~\eqref{E:grad:delta} is based on the facts that a modification in $\hat{\bbdelta}^{\ell}$ would reduce the cost $o(\bbomega)$ if: i) It more faithfully captures the empirical per-packet latency $\hat{\bbtau}^{\ell}$ (second term), and ii) Aggregated over jobs, the incorporation of the given link in the corresponding routes reduces $o(\bbomega)$ (first term).
Furthermore, recalling that $\hat{\bbdelta}^{\ell}=f(\ccalG^{\ell}, \bblambda^{\ell}, \bbr^{\ell}, \ccalG^c;\bbomega) $, expression \eqref{E:grad:omega} stems from applying the chain rule.
For each sampled instance $\varepsilon\sim\Omega$, $\bbomega$ is updated by stochastic gradient descent (SGD),  $\bbomega = \bbomega - \alpha \widehat{\nabla_{\bbomega} o(\bbomega)} $, 
% \red{SS: since we are doing gradient descent, shouldn't we move in the direction of the negative gradient?} 
where $\alpha>0$ is the learning rate.
The training ends based on an early stop mechanism.

\vspace{-0.1in}
\section{Numerical experiments}
\label{sec:results}
\vspace{-0.05in}

The GNN-enhanced distributed offloading is evaluated on simulated wireless ad-hoc networks. 
Each instance in the training and test sets of the form $\varepsilon=(\ccalG^{n}, \ccalG^{c}, \ccalM, \ccalR, \ccalS, \ccalJ, \bbr^c, \bbmu^n )$ is generated as follows.
The connectivity graph $ \ccalG^{n} $ is drawn as a random graph from the Barabási–Albert (BA)~\cite{Albert02} model, and the conflict graph $\ccalG^c$ is the line graph of $\ccalG^n$.
The BA model has two parameters: the number of vertices $|\ccalV|$ and the number of edges, $\nu $, that each new vertex forms during a preferential attachment process.
We set $\nu=2$, and $|\ccalV|\in\{20,30,\dots,110\}$.
The relay node set $\ccalR$ is selected as well-connected nodes by applying minimal node cut on  $ \ccalG^{n} $.
We then cut $\ccalV$ into a larger partition $ \ccalV^b$ and a smaller partition $\ccalV^s$ by solving the minimal cut problem on $\ccalG^n$ with the Stoer-Wagner algorithm~\cite{stoer1997simple}.
The server set $\ccalS$ contains $10\%\sim25\%$ of nodes, i.e., $|\ccalS|=\lfloor\mathbbm{U}(0.1,0.25)|\ccalV|\rceil$, randomly selected from $\ccalV^s-\ccalR$, and if $ |\ccalV^s-\ccalR| < |\ccalS| $ then from $ \ccalV^b-\ccalR$, where $ \mathbbm{U} $ stands for uniform distribution.
The rest are edge nodes, $\ccalM=\ccalV \setminus (\ccalR \cup \ccalS)$.
In this way, we make sure that server nodes are likely multiple hops away from edge nodes.
The link rate $\bbr^c_e\sim \mathbbm{U}(30,70),\; \forall\; e\in\ccalE$, and the service rate $\bbmu^n_v$ are drawn from Pareto distributions with shape $a=2$ and mode $q=100$ for $v\in\ccalS$ and $q=8$ for $v\in\ccalM$, and $ \bbmu^n_v=0, \forall\; v\in\ccalR $.
The training and test sets, respectively, contain $ 2000 $ and $1000$ network instances $ (\ccalG^{n}, \ccalG^{c}, \ccalM, \ccalR, \ccalS, \bbr^c, \bbmu^n ) $ generated under the same configuration but with different sets of pseudo-random seeds. 
The task set $\ccalJ$ is created randomly on-the-fly during training and testing, with job parameters $\eta^u(j_m)=100,\eta^d(j_m)=1$, $\lambda^j(j_m)\sim \mathbbm{U}(0.015,0.075)$, $\ccalS_m=\ccalS$, and $|\ccalJ|= \lfloor\mathbbm{U}(0.3,1)|\ccalM|\rceil $.
For each network instance, we draw 10 random instances of task set $\ccalJ$.
The scale of simulation instances by network size is illustrated in Fig.~\ref{fig:results:set}.

The hyperparameters of our GNN are $L=5, K=10, \alpha=10^{-6}$, and $g_l=32$ for $l\in\{1,2,3,4\}$.\footnote{Training takes $0.5$ hours on a workstation with a specification of 16GB memory, 8 cores, and Geforce GTX 1070 GPU. 
The source code is published at \url{https://github.com/zhongyuanzhao/multihop-offload}}
We limit the arrival of new jobs to the first $T=1000$ time slots, so that the execution latency of a task is always finite, even if it is congested.
A task $j_m\in\ccalJ$ is congested if the queues of any link $e\in\ccalE^e$ on its route in graph $ \ccalG^{\ell} $ is unstable, i.e., $ \hat{\bbmu}^{\ell}_e < \bbrho^{\ell}_e $ while evaluating $ \bbtau^{\ell} = \varphi(\ccalG^c, \ccalG^{\ell}, \bbr^{\ell}, \bbrho^{\ell}, T, K)$ (see line 10 of algorithm~\ref{algo:response}). 
If $j_m $ is congested, its latency $\bbu^j_{j_m} \geq T$.
The GNN-based offloading policy is compared against the baseline and local (all tasks computed locally without offloading) policies, on a set of $10000$ test instances, which is generated by creating 10 random task instances for each network instance in the test set. 

The average execution latency and task congestion ratio as a function of the network size under the tested policies are presented in Fig~\ref{fig:results:congest}.
The task congestion ratio is the ratio between the total number of congested tasks and the total number of tasks for the 10 random task instances on each network instance.
The average execution latency under the baseline policy is $288.7$ compared to the GNN-based ($18.4$) and local ($20.4$) policies due to its high congestion ratio (up to $14.5\%$ in larger networks).
Both the GNN-based and local policies can avoid task congestion under the test traffic configuration, whereas the GNN-based policy has the lowest execution latency across different network sizes, which on average is $9.8\%$ lower than the local policy. 

In Fig.~\ref{fig:results:response}, we present the boxplot of the average per-task latency ratio of the GNN and local policies w.r.t. the baseline across network sizes, defined as $ \mathbbm{E}_{\ccalJ}(\hat{\bbu}^{j}_{j_m}/\bar{\bbu}^{j}_{j_m}) $ per test instance.
This metric can better describe the effects of a policy on free-flowing tasks under the baseline, as it is not dominated by the congested tasks like the average execution latency, e.g., $ \hat{\bbu}^{j}_{j_m}/\bar{\bbu}^{j}_{j_m} \approx 0 $ if task $j_m$ is congested under the baseline.
The average per-task latency ratios of GNN-based and local policies are respectively $1.91$ and $2.24$. 
It shows that the GNN-based policy is in between the most conservative local policy that prevents congestion by avoiding offloading altogether, and the baseline policy that offloads tasks aggressively for faster execution but often causes congestion in the network.

% \begin{figure}
%     \centering
%     \includegraphics[width=0.8\linewidth]{figures/Adhoc_Adhoc_test_data_aco_data_ba_200_load_0_10_training_trajectory.pdf}
%     \vspace{-0.1in}
%     \caption{Trajectory of system objective $o(\bbomega)=\boldsymbol{1}^{\top}\hat{\bbu}^j$ in training. \textcolor{blue}{GV: If space is tight, I propose removing this figure and replacing it with a table of notation used in the paper and/or extra text to explain the main ideas behind the design choices of the algorithm more, e.g. the intuition underlying Algorithm 1.}}
%     \label{fig:training}
%     \vspace{-0.1in}
% \end{figure}

\vspace{-0.05in}
\section{Conclusions}
\label{sec:conclusions}
\vspace{-0.05in}
In this paper, we develop a fully-distributed approach for computational offloading in wireless multi-hop networks.
The key idea is to use graph neural networks to encode the information of computational tasks, links, and servers across the network into link weights, which can bring the awareness of other tasks across the network to distributed offloading and routing decisions that are otherwise agnostic to that information.
In this way, we can lower the execution latency of tasks by reducing the congestion in the network. 
% By reducing task congestion, the GNN-based offloading policy can support more heavy network traffic than the baseline approach which only applies to ultra lightweight traffic.
Compared with traditional methods, where each mobile device estimates the costs of its offloading options through probing packets, our approach can better understand the network context at lower communication overhead.
Future work includes improving the training method and GNN design for better latency and energy performance. 
\vfill\pagebreak

%\section{REFERENCES}
%\label{sec:refs}

% -------------------------------------------------------------------------

\bibliographystyle{ieeetr}
\bibliography{strings,refs}

\begin{thebibliography}{10}

\bibitem{sarkar2013ad}
S.~K. Sarkar, T.~G. Basavaraju, and C.~Puttamadappa, {\em Ad hoc Mobile
  Wireless Networks: Principles, Protocols and Applications}.
\newblock CRC Press, 2013.

\bibitem{kreutz2014software}
D.~Kreutz, F.~M. Ramos, P.~E. Verissimo, C.~E. Rothenberg, S.~Azodolmolky, and
  S.~Uhlig, ``Software-defined networking: A comprehensive survey,'' {\em
  Proceedings of the IEEE}, vol.~103, no.~1, pp.~14--76, 2014.

\bibitem{kott2016internet}
A.~Kott, A.~Swami, and B.~J. West, ``The internet of battle things,'' {\em
  Computer}, vol.~49, no.~12, pp.~70--75, 2016.

\bibitem{cisco2020}
``Cisco annual internet report (2018–2023),'' white paper, Cisco Systems,
  Inc., Mar. 2020.

\bibitem{akyildiz20206g}
I.~F. Akyildiz, A.~Kak, and S.~Nie, ``{6G} and beyond: The future of wireless
  communications systems,'' {\em IEEE Access}, vol.~8, pp.~133995--134030,
  2020.

\bibitem{chen2021massive}
X.~Chen, D.~W.~K. Ng, W.~Yu, E.~G. Larsson, N.~Al-Dhahir, and R.~Schober,
  ``Massive access for {5G} and beyond,'' {\em IEEE J. Sel. Areas Commun.},
  vol.~39, no.~3, pp.~615--637, 2021.

\bibitem{noor20226g}
M.~Noor-A-Rahim, Z.~Liu, H.~Lee, M.~O. Khyam, J.~He, D.~Pesch, K.~Moessner,
  W.~Saad, and H.~V. Poor, ``6g for vehicle-to-everything (v2x) communications:
  Enabling technologies, challenges, and opportunities,'' {\em Proceedings of
  the IEEE}, vol.~110, no.~6, pp.~712--734, 2022.

\bibitem{tassiulas1992}
L.~Tassiulas, ``Stability properties of constrained queueing systems and
  scheduling policies for maximum throughput in multihop radio networks,'' {\em
  IEEE Trans. on Automatic Control}, vol.~31, no.~12, 1992.

\bibitem{Joo09}
C.~{Joo}, X.~{Lin}, and N.~B. {Shroff}, ``Understanding the capacity region of
  the greedy maximal scheduling algorithm in multihop wireless networks,'' {\em
  IEEE/ACM Trans. Netw.}, vol.~17, no.~4, pp.~1132--1145, 2009.

\bibitem{Lin2009constant}
X.~Lin and S.~B. Rasool, ``Constant-time distributed scheduling policies for ad
  hoc wireless networks,'' {\em IEEE Trans. on Automatic Control}, vol.~54,
  no.~2, pp.~231--242, 2009.

\bibitem{jiang2010distCSMA}
L.~Jiang and J.~Walrand, ``A distributed {CSMA} algorithm for throughput and
  utility maximization in wireless networks,'' {\em IEEE/ACM Trans. Netw.},
  vol.~18, no.~3, pp.~960--972, 2010.

\bibitem{zhao2022twc}
Z.~Zhao, G.~Verma, C.~Rao, A.~Swami, and S.~Segarra, ``Link scheduling using
  graph neural networks,'' {\em IEEE Trans. Wireless Commun.}, vol.~22, no.~6,
  pp.~3997--4012, 2023.

\bibitem{zhao2023delay}
Z.~Zhao, B.~Radojicic, G.~Verma, A.~Swami, and S.~Segarra, ``Delay-aware
  backpressure routing using graph neural networks,'' in {\em IEEE Int. Conf.
  on Acoustics, Speech and Signal Process. (ICASSP)}, pp.~4720--4724, 2023.

\bibitem{zhao2023graphbased}
Z.~Zhao, A.~Swami, and S.~Segarra, ``Graph-based deterministic policy gradient
  for repetitive combinatorial optimization problems,'' in {\em Intl. Conf.
  Learn. Repres. (ICLR)}, 2023.

\bibitem{van2018quality}
D.~Van~Le and C.-K. Tham, ``Quality of service aware computation offloading in
  an ad-hoc mobile cloud,'' {\em IEEE Trans. Vehicular Tech.}, vol.~67, no.~9,
  pp.~8890--8904, 2018.

\bibitem{ferrer2019towards}
A.~J. Ferrer, J.~M. Marqu{\`e}s, and J.~Jorba, ``Towards the decentralised
  cloud: Survey on approaches and challenges for mobile, ad hoc, and edge
  computing,'' {\em ACM Computing Surveys (CSUR)}, vol.~51, no.~6, pp.~1--36,
  2019.

\bibitem{funai2019computational}
C.~Funai, C.~Tapparello, and W.~Heinzelman, ``Computational offloading for
  energy constrained devices in multi-hop cooperative networks,'' {\em IEEE
  Transactions on Mobile Computing}, vol.~19, no.~1, pp.~60--73, 2019.

\bibitem{chattopadhyay2020fully}
R.~Chattopadhyay and C.-K. Tham, ``Fully and partially distributed incentive
  mechanism for a mobile edge computing network,'' {\em IEEE Transactions on
  Mobile Computing}, vol.~21, no.~1, pp.~139--153, 2020.

\bibitem{cai2020mobile}
Y.~Cai, J.~Llorca, A.~M. Tulino, and A.~F. Molisch, ``Mobile edge computing
  network control: Tradeoff between delay and cost,'' in {\em IEEE Global
  Communications Conference (GLOBECOM)}, pp.~1--6, 2020.

\bibitem{feng2021multi}
G.~Feng, X.~Li, Z.~Gao, C.~Wang, H.~Lv, and Q.~Zhao, ``Multi-path and multi-hop
  task offloading in mobile ad hoc networks,'' {\em IEEE Trans. Vehicular
  Tech.}, vol.~70, no.~6, pp.~5347--5361, 2021.

\bibitem{liu2022mobility}
L.~Liu, M.~Zhao, M.~Yu, M.~A. Jan, D.~Lan, and A.~Taherkordi, ``Mobility-aware
  multi-hop task offloading for autonomous driving in vehicular edge computing
  and networks,'' {\em IEEE Transactions on Intelligent Transportation
  Systems}, vol.~24, no.~2, pp.~2169--2182, 2022.

\bibitem{dai2022learning}
X.~Dai, Z.~Xiao, H.~Jiang, H.~Chen, G.~Min, S.~Dustdar, and J.~Cao, ``A
  learning-based approach for vehicle-to-vehicle computation offloading,'' {\em
  IEEE Internet of Things Journal}, vol.~10, no.~8, pp.~7244--7258, 2022.

\bibitem{li2022novel}
X.~Li, T.~Chen, D.~Yuan, J.~Xu, and X.~Liu, ``A novel graph-based computation
  offloading strategy for workflow applications in mobile edge computing,''
  {\em IEEE Transactions on Services Computing}, vol.~16, no.~2, pp.~845--857,
  2022.

\bibitem{cheng2009complexity}
W.~Cheng, X.~Cheng, T.~Znati, X.~Lu, and Z.~Lu, ``The complexity of channel
  scheduling in multi-radio multi-channel wireless networks,'' in {\em IEEE
  Intl. Conf. on Computer Comms. (INFOCOM)}, pp.~1512--1520, 2009.

\bibitem{yang2016learning}
J.~Yang, S.~C. Draper, and R.~Nowak, ``Learning the interference graph of a
  wireless network,'' {\em IEEE Trans. Signal Inf. Process. Netw.}, vol.~3,
  no.~3, pp.~631--646, 2016.

\bibitem{little1961proof}
J.~D. Little, ``A proof for the queuing formula: L= $\lambda$ w,'' {\em
  Operations research}, vol.~9, no.~3, pp.~383--387, 1961.

\bibitem{oncan2007survey}
T.~{\"O}ncan, ``A survey of the generalized assignment problem and its
  applications,'' {\em INFOR: Information Systems and Operational Research},
  vol.~45, no.~3, pp.~123--141, 2007.

\bibitem{harary1960some}
F.~Harary and R.~Z. Norman, ``Some properties of line digraphs,'' {\em
  Rendiconti del circolo matematico di palermo}, vol.~9, pp.~161--168, 1960.

\bibitem{bellman1958routing}
R.~Bellman, ``On a routing problem,'' {\em Quarterly of Applied Mathematics},
  vol.~16, no.~1, pp.~87--90, 1958.

\bibitem{ford1956network}
L.~R. Ford~Jr, ``Network flow theory,'' tech. rep., Rand Corp Santa Monica CA,
  1956.

\bibitem{bernstein2019distributed}
A.~Bernstein and D.~Nanongkai, ``Distributed exact weighted all-pairs shortest
  paths in near-linear time,'' in {\em ACM SIGACT Symp. Theory of Comp.},
  pp.~334--342, 2019.

\bibitem{Albert02}
R.~Albert and A.-L. Barab\'asi, ``Statistical mechanics of complex networks,''
  {\em Rev. Mod. Phys.}, vol.~74, pp.~47--97, Jan 2002.

\bibitem{stoer1997simple}
M.~Stoer and F.~Wagner, ``A simple min-cut algorithm,'' {\em Journal of the ACM
  (JACM)}, vol.~44, no.~4, pp.~585--591, 1997.

\end{thebibliography}

\end{document}